\documentclass[10pt,letterpaper]{article}
\usepackage[top=0.85in,left=2.75in,footskip=0.75in]{geometry}

\usepackage{amsmath,amssymb}

\usepackage{changepage}

\usepackage[utf8x]{inputenc}

\usepackage{textcomp,marvosym}

\usepackage{cite}

\usepackage{nameref,hyperref}

\usepackage[right]{lineno}

\usepackage{microtype}
\DisableLigatures[f]{encoding = *, family = * }

\usepackage[table]{xcolor}

\usepackage{array}

\newcolumntype{+}{!{\vrule width 2pt}}

\usepackage{multirow}

\newlength\savedwidth

\newcommand\thickhline{\noalign{\global\savedwidth\arrayrulewidth\global\arrayrulewidth 2pt}%
\hline
\noalign{\global\arrayrulewidth\savedwidth}}

\raggedright
\usepackage[aboveskip=1pt,labelfont=bf,labelsep=period,justification=raggedright,singlelinecheck=off]{caption}

\makeatletter
\renewcommand{\@biblabel}[1]{\quad#1.}
\makeatother

\date{}

\usepackage{lastpage,fancyhdr,graphicx}
\usepackage{epstopdf}
\fancyheadoffset[L]{2.25in}
\fancyfootoffset[L]{2.25in}
\begin{document}
\vspace*{0.2in}

\begin{flushleft}
{\Large
\textbf\newline{Strategy intervention for the evolution of fairness} 
}
\newline
\\
Yanling Zhang\textsuperscript{1},
Feng Fu\textsuperscript{2,3*},
\\
\bigskip
\textbf{1} Key Laboratory of Knowledge Automation for Industrial Processes of Ministry of Education, School of Automation and Electrical Engineering, University of Science and Technology Beijing, Beijing, China\\
\textbf{2} Department of Mathematics, Dartmouth College,
Hanover, New Hampshire, United States of America
\\
\textbf{3} Department of Biomedical Data Science, Geisel
School of Medicine, Dartmouth College, Hanover, New Hampshire, United States of America
\\
\bigskip

%
%






* fufeng@gmail.com (FF)

\end{flushleft}
\section*{Abstract}
Masses of experiments have shown individual preference for fairness which seems irrational.
The reason behind it remains a focus for research.
The effect of spite (individuals are only concerned with their own relative standing) on the evolution of fairness has attracted increasing attention from experiments, but only has been implicitly studied in one evolutionary model.
The model did not involve high-offer rejections, which have been found in the form of  non-monotonic rejections (rejecting offers that are too high or too low) in experiments.
Here, we introduce a high offer and a non-monotonic rejection in structured populations of finite size, and use strategy intervention to explicitly study how spite influences the evolution of
fairness: five strategies are in sequence added into the competition of a fair strategy and a selfish strategy.
We find that spite promotes fairness, altruism inhibits fairness, and the non-monotonic rejection can cause fairness to overcome selfishness, which cannot happen without high-offer rejections.
Particularly for the group-structured population with seven discrete strategies, we analytically study the effect of population size, mutation, and migration on fairness, selfishness, altruism, and spite. A larger population size cannot change the dominance of fairness, but it promotes altruism and inhibits selfishness and spite.
Intermediate mutation maximizes selfishness and fairness, and minimizes spite; intermediate mutation maximizes altruism for intermediate migration and minimizes altruism otherwise.
The existence of migration inhibits selfishness and fairness, and promotes altruism; sufficient migration promotes spite.
Our study may provide important insights into the evolutionary origin of fairness.

\section*{Author summary}
In game theory, the traditional assumption that players are rational (motivated purely by payoff consideration) predicts that individuals are selfish.
In contrast to this prediction, masses of Ultimatum Game experiments have showed individual preference for fairness.
To expose the mechanisms behind fairness, many evolutionary game models have been proposed.
One of them implicitly studied the effect of spite (only concerns with one's own relative standing) on fairness by the competition of four discrete strategies which represent selfishness, fairness, altruism, and spite, respectively.
The model did not involve high-offer rejections, which have been found in the form of  non-monotonic rejections (rejecting high and low offers) in experiments.
Here, we introduce a high offer and a non-monotonic rejection to form seven discrete strategies besides the above four strategies, and three new strategies represent fairness, altruism, and spite, respectively.
Moreover, strategy intervention is used to explicitly study how spite influences
fairness: we add five strategies to the competition between a selfish strategy and a fair strategy in sequence, and our study goes from the three-strategy competition to the seven-strategy competition.
For group-structured populations with the seven discrete strategies, we analytically study how population size, mutation, and migration influence selfishness, fairness, altruism, and spite.

\section*{Introduction}
For over thirty years, there has been substantial progress in understanding the evolution of fairness based on the Ultimatum Game (UG).
In the UG, a proposer suggests a way to allocate a fixed sum of money between herself and a responder, and then the responder decides whether to accept the allocation method or not.
If the allocation method is accepted by the responder, both of them are paid accordingly; if rejected, neither of them gets paid.
If an individual only cares about maximizing his own payoff (which leads to the subgame perfect equilibrium), the responder should accept any non-zero offers, and thus the proposer offers the minimum allowable proportion of the sum to the responder.
However, this prediction contradicts with almost all experimental observations, in which the most common allocation method is the half-half split and the offers less than $30\%$ of the sum are frequently rejected (see reviews~\cite{Thaler1988,Guth2014}).

In most UG experiments, individuals are observed to only reject low offers (less than $50\%$ of the sum).
However, some UG experiments~\cite{Huck1999,Bahry2006,Henrich2001,Hennig-Schmidt2008} have found the rejection of high offers (greater than $50\%$ of the sum) in \emph{nonstudent} populations.
It has been suggested that the reason why most previous experiments did not show high offer rejections is that they are confined to typical student subject pools~\cite{Henrich2006}.
The rejection of an offer can be regarded as the behavior of costly punishment~\cite{Rand2011,Perc2015,Bauch2016} because both the responder and the proposer, who would gain payoffs if the offer were accepted, receive nothing and suffer the corresponding cost.
Some investigations argue that such rejection is motivated by individual prosocial preference for fairness and indicates the existence of `strong reciprocity' (a behavioral propensity to voluntarily cooperate, if treated fairly, and punish non-cooperators)~\cite{Fehr2002,Dawes2007}.
However, the preference for fairness is not the only motivation behind rejecting an offer.
By comparing individuals' propensity to reject unfair offers in the UG with their tendency to perform various prosocial behaviors in other games, it has been found that the preference for spite is another potential motivation~\cite{Yamagishi2012,Garza2014}.
This finding suggests that we should pay attention to the evolution of spite and its role in the evolution of fairness to explain the UG experimental behaviors.

The evolution of spite and its role in the evolution of fairness have attracted increasing attention from experiments, but only have been studied by one recent evolutionary model~\cite{Forber2014}.
This model used
a simplified version of the UG with four discrete strategies $(\mbox{low}, \mbox{accept any})$, $(\mbox{fair}, \mbox{reject low})$, $(\mbox{fair}, \mbox{accept any})$, and $(\mbox{low}, \mbox{reject low})$, which represent selfishness, fairness, altruism, and spite, respectively.
It did not involve high-offer rejections, which have been found in the form of non-monotonic rejections (rejecting offers that are too high and too low) in experiments~\cite{Bahry2006,Hennig-Schmidt2008}.
In this paper, we will introduce a high offer and a non-monotonic rejection to form seven discrete strategies besides the above four strategies, and three new strategies represent altruism, spite, and fairness, respectively.
The previous model~\cite{Forber2014} has implicitly given the effect of spite on the evolution of fairness by analyzing the competition of four discrete strategies.
Unlike the previous model, strategy intervention will be used here: we will first study the competition between a selfish strategy and a fair strategy; we will then add five strategies to them in sequence, and our study will go from the three-strategy competition to the seven-strategy competition. Accordingly, we can explicitly investigate the effect of spite and altruism on the evolution of fairness.

In this paper, we will focus on the evolution of fairness in finite populations, and yet the above-mentioned previous model~\cite{Forber2014} assumed infinitely large populations.
It is worth mentioning that our study with four strategies recovers to the previous model but with finite populations.
The evolution of fairness has been widely studied by the evolutionary dynamics of the UG~\cite{Nowak2000,Page2002,Page2000,Killingback2001,Iranzo2011,Szolnoki2012a,Szolnoki2012b,Wu2013,Wang2015,
Ichinose2014,Rand2013, Zhang2013}, in which strategies with higher fitness are more likely to spread among populations.
Evolutionary dynamics~\cite{masuda09,masuda15,Bauch2012,Bonhoeffer2015,Bonhoeffer2016,Beerenwinkel2016,Perc2017}
 could characterize genetic evolution and cultural evolution, both of which have been used to account for the UG experimental behavior~\cite{Wallace2007,Henrich2001}.
The classic approach to evolutionary game dynamics is deterministic, and holds for infinitely large well-mixed populations.
Such deterministic dynamics shows that fairness cannot evolve without additional mechanisms~\cite{Nowak2000}.
To promote the evolution of fairness under the evolutionary framework, many additional mechanisms have been proposed: reputation (the proposer knows what offers the responder has accepted in the past)~\cite{Nowak2000}, empathy (individuals make offers which they would be prepared to accept)~\cite{Page2002}, spatial structures~\cite{Page2000,Killingback2001,Iranzo2011,Szolnoki2012a,Szolnoki2012b,Wu2013,Wang2015}, and repeated interactions~\cite{Ichinose2014}.
Without these additional mechanisms, fairness has also been found to evolve in finite populations even with the well-mixed structure~\cite{Rand2013}, suggesting that randomness plays a vital role in the evolution of fairness.

In this paper, we will study structured populations which satisfy the well-known Tarnita-$\sigma$ condition~\cite{Tarnita2011}.
The Tarnita-$\sigma$ condition is that strategy $k\in\{1,2,\cdots, S\}$ is favored by natural selection (the average frequency of strategy $k$ over the stationary distribution is greater than $1/S$) under weak selection if and only if
\begin{eqnarray}
\begin{array}{l}
\Gamma_1(a_{kk}-\overline{a}_{**})+\Gamma_2
(\overline{a}_{k*}-\overline{a}_{*k})+\Gamma_3(\overline{a}_{k*}-\overline{a})>0
\end{array}
\label{xcc}
\end{eqnarray}
where $a_{ij}$ is the payoff of an individual using strategy $i$ when interacting with an individual using strategy $j$, $\overline{a}_{**}=\frac{1}{S}\sum_{i=1}^S a_{ii}$, $\overline{a}_{k*}=\frac{1}{S}\sum_{i=1}^S a_{ki}$, $\overline{a}_{*k}=\frac{1}{S}\sum_{i=1}^S a_{ik}$, and $\overline{a}=\frac{1}{S^2}\sum_{i=1}^S\sum_{j=1}^S a_{ij}$.
Two parameters $\sigma_1=\frac{\Gamma_1}{\Gamma_2}$ and $\sigma_2=\frac{\Gamma_3}{\Gamma_2}$ quantify the dependence of the multi-strategy selection on the pairwise competition and the competition of all strategies with equal frequency, respectively.
The Tarnita-$\sigma$ condition holds for a large class of finite populations in which the population structure and the update rule satisfy some mild assumptions.
For example, the population structure could involve interactions between neighbor nodes on a graph~\cite{Chen2014} or between individuals of the same group, phenotype, or set~\cite{Antal2009B,Tarnita2009B,Fu2013}, and the update rule could be the Moran process, the Wright-Fisher process, or the pairwise comparison process.
For structured populations satisfying the Tarnita-$\sigma$ condition, we will first separately investigate the impact of altruism and spite on the competition of selfishness and fairness, which has not been explicitly studied so far.
The unknown parameters $\Gamma_1, \Gamma_2,$ and $\Gamma_3$ in equation~(\ref{xcc}) are difficult to calculate for general models.
However for group-structured populations together with the Moran process or the Wright-Fisher process, they can be calculated based on the results in the prior literature~\cite{Zhang2016b}.
After calculating $\Gamma_1, \Gamma_2,$ and $\Gamma_3$ in Table~\ref{table3}, we will use them to quantitatively analyze the effect of population size, mutation, or migration on selfishness, fairness, altruism, and spite,  and compare the effects from two update rules.
A group of group-structured populations can be understood as an island in population genetics or a particular company in human society.
The group-structured population with one group is the well-mixed population.
The long-term group-structured population without migration evolves just like the well-mixed population: the absorbing state has all individuals in one group.
Accordingly in the absence of migration, the results are appropriate for the well-mixed population.

\section*{Results}
\subsection*{Structured populations satisfying the Tarnita-$\sigma$ condition}
\begin{figure}[!h]
\includegraphics[width=\linewidth]{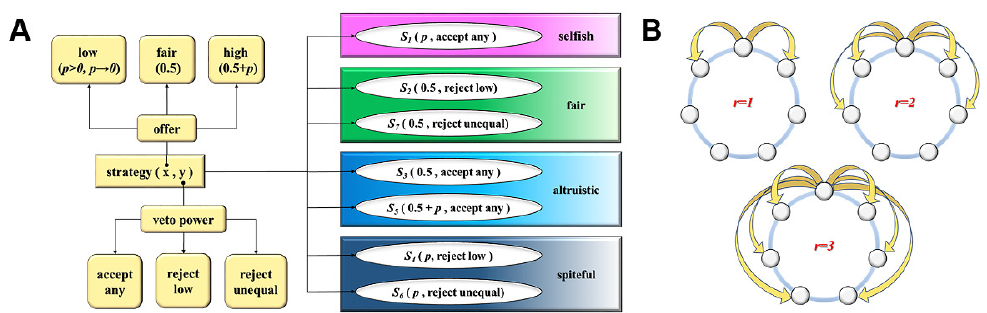}
\caption{{\bf Evolutionary dynamics in structured populations satisfying the Tarnita-$\sigma$ condition}.
A: The competition of selfishness ($S_1$) and fairness ($S_2$ and $S_7$) is influenced by altruism ($S_3$ and $S_5$) and spite ($S_4$ and $S_6$).
B: For the seven-strategy competition, the scores of $S_1$, $S_2$, $S_3$, $S_4$, $S_5$, $S_6$, and $S_7$ are calculated.
When a strategy $X$ offers more to one other strategy $Y$ ($Y\neq X$), an arrow starts from $X$ and ends at $Y$, and then $X$ loses one score and $Y$ obtains one score.}
\label{fig1}
\end{figure}

When the selfish strategy $S_1$ and the fair strategy with the low-offer rejection $S_2$ coexist in the population without other strategies, they compete equally with each other, i.e., $f_1=f_2$, because their payoffs are identical for any population states.
Fig~\ref{fig1} demonstrates how the competition between selfishness and fairness is influenced by altruism and spite through adding $S_3$, $S_4$, $S_5$, $S_6$, and $S_7$ in sequence.
When the altruistic strategy with the fair offer $S_3$ is introduced into the population with $S_1$ and $S_2$,
$S_1$ gains an advantage over $S_2$, i.e., $f_1>f_2$.
This advantage could be offset, i.e., $f_1=f_2$, after the further addition of the spiteful strategy with the low-offer rejection $S_4$.
Similar to $S_3$, the altruistic strategy with the high offer $S_5$ induces $S_1$ to gain an advantage over $S_2$ again ($f_1>f_2$) in the five-strategy competition.
Similar to $S_4$, the spiteful strategy with the unequal-offer rejection $S_6$ erases this advantage ($f_1=f_2$) in the six-strategy competition.
The advantage of $S_1$ over $S_2$ appears because the altruistic strategies $S_3$ and $S_5$ are exploited more severely by $S_1$ than $S_2$,
and disappears because the spiteful strategies $S_4$ and $S_6$ punish $S_1$ more severely than $S_2$.
This can be understood intuitively by comparing the row sum of $S_1$ with that of $S_2$ in
Table~\ref{table1}.
Similarly, it can be seen when the fair strategy with the unequal-offer rejection $S_7$ is introduced, the advantage of $S_2$ over $S_1$ is greater than the advantage of $S_1$ over $S_7$.
Therefore, fairness ($S_2$ and $S_7$) first gains an advantage over selfishness ($S_1$), i.e., $f_2>f_1$, in the seven-strategy competition.
This means that the non-monotonic rejection can cause fairness to overcome selfishness, which cannot happen without the high-offer rejection.

For the seven-strategy competition, $f_1$, $f_2$, and $f_3$ increase with $p$, and $f_4$ decreases with $p$ (see Table~\ref{table2}).
The conclusion can be intuitively understood from Table~\ref{table1}.
Specifically, we have the following conclusions about $S_1$, $S_4$, $S_5$, and $S_6$ when $p$ increases: the selfish strategy $S_1$ offers more to $S_3$ and $S_5$, and its offer is rejected by the remaining other strategies;
the spiteful strategies $S_4$ and $S_6$ offer more to $S_1$, $S_3$, and $S_5$, and their offers are rejected by the remaining other strategies;
the altruistic strategy $S_5$ offers more to $S_1$, $S_2$, $S_3$, and $S_4$, and its offer is rejected by the remaining other strategies.
When a strategy $X$ offers more to one other strategy $Y$ ($Y\neq X$),
we assume that $X$ loses one score and $Y$ obtains one score.
After all possible effects of $S_1$, $S_4$, $S_5$, and $S_6$ on other strategies are considered, the total score of each strategy determines how the corresponding $f$ changes with $p$: $f$ increases or decreases with $p$ if the total score is positive or negative, and $f$ does not change with $p$ if the total score is zero.
For the seven-strategy competition (Fig~\ref{fig1}),
$S_1$ scores $1$, the total score of $S_2$ and $S_7$ is $1$, the total score of $S_3$ and $S_5$ is $3$, and the total score of $S_4$ and $S_6$ is $-5$.
Therefore, $f_1$, $f_2$, and $f_3$ increase with $p$, and $f_4$ decreases with $p$.
Similarly, we can intuitively understand the change of $f_i, i\in\{1,2,3,4\}$ with $p$ from the three-strategy competition to the six-strategy competition.
For the rest of the paper, we will only focus on the case of $p=0.01$.

\subsection*{Group-structured populations}
\begin{figure}[h!]
\includegraphics[width=\linewidth]{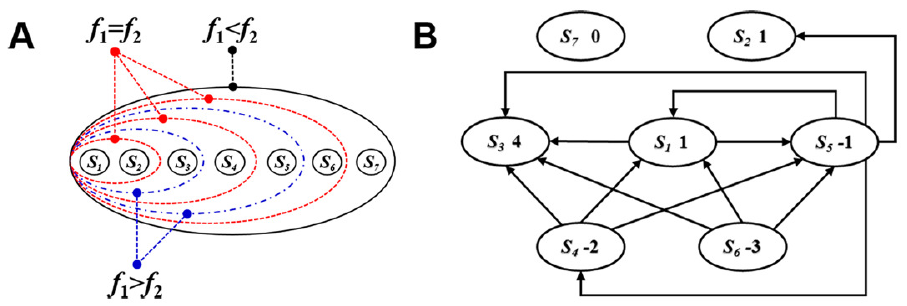}
\caption{{\bf Comparison of theoretical results and simulated results in the Moran process (first row) and the Wright-Fisher process (second row).}
For the frequencies of selfishness (first column), fairness (second column), altruism (third column), and spite (fourth column), the difference of the theoretical values and the simulated values (averaged over $5\times10^8-10^6$ generations) is negligible when the selection intensity $\delta$ is sufficiently small, and is no longer negligible for other $\delta$.
Parameters: $N=50$, $M=7$, $u=0.1$, $v=0.1$, $r=1$, and $p=0.01$.}
\label{fig2}
\end{figure}

Hereafter, the seven-strategy competition will be investigated in group-structured populations. For the Moran process and the Wright-Fisher process,
the average frequency of strategy $k\in\{1,2,\cdots, S\}$ over the stationary distribution under weak selection $(\delta\to0)$, $\langle x_k\rangle_{\delta\to 0}$, is given by
\begin{eqnarray}
\label{xcc1}
\begin{array}{l}
\langle x_{k}\rangle_{\delta\rightarrow 0}
=\frac{1}{S}+\delta\frac{1-u}{Nu}(\Gamma_1(a_{kk}-\overline{a}_{**})+\Gamma_2
(\overline{a}_{k*}-\overline{a}_{*k})+\Gamma_3(\overline{a}_{k*}-\overline{a}))
\end{array}
\end{eqnarray}
in which $\Gamma_1$, $\Gamma_2$, and $\Gamma_3$ are unknown.
Here, $\Gamma_1$, $\Gamma_2$, and $\Gamma_3$ can be expressed by
$\langle x_iI_{jk}\rangle_0$ which is the probability-weighted average of $x_iI_{jk}$
over all possible steady states under neutral selection (the subscript $0$):
\begin{eqnarray}
\label{gamma}
\begin{array}{l}
\Gamma_1=S(\langle x_1I_{22}\rangle_0-\langle x_1I_{23}\rangle_0)\quad\Gamma_2=S(\langle x_1I_{12}\rangle_0-\langle x_1I_{23}\rangle_0)\quad\Gamma_3=S^2\langle x_1I_{23}\rangle_0
\end{array}
\end{eqnarray}
where $I_{ij}$ is the total number of interactions between individuals using strategy $i$ and individuals using strategy $j$ (each interaction between two individuals using strategy $i$ is counted twice in computing $I_{ii}$).
The precise value of $\langle x_iI_{jk}\rangle_0$ has been calculated for the Moran process and the Wright-Fisher process in the prior literature~\cite{Zhang2016b}.
By using the known value of $\langle x_iI_{jk}\rangle_0$, we obtain the precise expressions of $\Gamma_1$, $\Gamma_2$, and $\Gamma_3$, which are shown in Table~\ref{table3}.
These expressions hold for any population sizes, non-zero mutation probabilities, migration probabilities, migration ranges, and group numbers.
Migration of the range $r$ is characterized by $f(x)$, and it is given as
\begin{equation}
\label{fx}
\begin{array}{ll}
f(x)=\frac{1}{M-1}\sum_{j=1}^{M-1}\cos\frac{2\pi j x}{M}\quad \mbox{if $r=\frac{M}{2}$ for even $M$} \\
f(x)=\frac{1}{r}\sum_{j=1}^{r}\cos\frac{2\pi j x}{M} \quad\mbox{else if $r=1,2,\cdots, \lfloor \frac{M}{2}\rfloor$} \\
\end{array}
\end{equation}
where $\lfloor \frac{M}{2}\rfloor$ is the greatest integer no greater than $\frac{M}{2}$.
Fig~\ref{fig2} shows that the difference of theoretical results and simulated results is negligible when the selection intensity $\delta$ is sufficiently small and is no longer negligible for other $\delta$. We will study how the population size $N$, the mutation probability $u$, the migration probability $v$, and the migration range $r$ influence the evolution of selfishness, fairness, altruism, and spite.

\begin{table}[!ht]
\begin{adjustwidth}{-2.25in}{0in} 
\centering
\caption{
{\boldmath $\Gamma_1$, $\Gamma_2$} {\bf and} {\boldmath $\Gamma_3$} {\bf for the Moran process} ({\boldmath $\Gamma_1^{Mo}$, $\Gamma_2^{Mo}$,} {\bf and} {\boldmath $\Gamma_3^{Mo}$} {\bf respectively}),
{\bf and} {\boldmath $\Gamma_1$, $\Gamma_2$} {\bf and} {\boldmath $\Gamma_3$} {\bf for the Wright-Fisher process} ({\boldmath $\Gamma_1^{WF}$, $\Gamma_2^{WF}$,} {\bf and} {\boldmath $\Gamma_3^{WF}$} {\bf respectively}), {\bf where}
{\boldmath $\Phi_i(f(x))$, $\Psi_i(f(x))$, $\Phi_i^{'}(f(x))$,} {\bf and} {\boldmath $\Psi_i^{'}(f(x))$} {\bf are abbreviated as} {\boldmath $\Phi_i$, $\Psi_i$$, \Phi_i^{'}$,} {\bf and} {\boldmath $\Psi_i^{'}$}.}
\begin{tabular}{|l+l|l|}
\hline
{\boldmath $\Gamma_1^{Mo}$} &$(N-1)(N-2)/(3M)\sum_{x=1}^M(-2\Phi_1\Psi_2-\Phi_4\alpha_1+3\Psi_2)$\\\hline
{\boldmath $\Gamma_2^{Mo}$}
&$(N-1)/(3M)\sum_{x=1}^M(3\Psi_1-3\Psi_2 +(N-2)(-2\Phi_1\Psi_2-\Phi_4\alpha_1+\Phi_2\Psi_2+\Phi_3\Psi_1+\Phi_5\alpha_1))$\\\hline
{\boldmath $\Gamma_3^{Mo}$}
&$(N-1)(N-2)/(3M)\sum_{x=1}^M(3\Psi_1-3\Psi_2+2(2\Phi_1\Psi_2+\Phi_4\alpha_1-\Phi_2\Psi_2-\Phi_3\Psi_1-\Phi_5\alpha_1))$\\\hline
{\boldmath $\Gamma_1^{WF}$}
&$(N-1)(N-2)/(3M)\sum_{x=1}^M(-\Phi_2^{'}(2N\Psi_2^{'}+N\alpha_1^{'}-2)+\Phi_3^{'}(2N\Psi_2^{'}+N-2))$\\\hline
{\boldmath $\Gamma_2^{WF}$}&$ (N-1)/(3M)\sum_{x=1}^M(\Phi_1^{'}(2N\Psi_1^{'}+N-2)-(N-2)\Phi_2^{'}(2N\Psi_2^{'}+N\alpha_1^{'}-2)$\\
&$-\Phi_3^{'}(2N\Psi_2^{'}+N-2)
+(N-2)\Phi_3^{'}(N\Psi_2^{'}+N\Psi_1^{'}+N\alpha_1^{'}-2))$\\\hline
{\boldmath $\Gamma_3^{WF}$}&$(N-1)(N-2)/(3M)\sum_{x=1}^M(\Phi_1^{'}(2N\Psi_1^{'}+N-2)+2\Phi_2^{'}(2N\Psi_2^{'}+N\alpha_1^{'}-2)$\\
 &$-\Phi_3^{'}(2N\Psi_2^{'}+N-2)
-2\Phi_3^{'}(N\Psi_2^{'}+N\Psi_1^{'}+N\alpha_1^{'}-2))$\\
\hline\end{tabular}
\begin{flushleft}
\begin{scriptsize}$\alpha_1=\frac{1-u}{1+(N-1)u}$\,
$\Phi_1(f)=\frac{(1-u)(2-v(1-f))}{2+(N-2)u+\frac{2(N-2)(1-u)v}{3}(1-f)}$\, $\Phi_2(f)=\frac{2-u-v(1-f)}{2+\frac{2(N-2)u}{3}+\frac{(N-2)(2-u)v}{3}(1-f)}$\,$\Phi_3(f)=\frac{(1-u)(2-v(1-f))}{2+\frac{2(N-2)u}{3}+\frac{(N-2)(2-u)v}{3}(1-f)}$\end{scriptsize}\\
\begin{scriptsize}$\Phi_4(f)=\frac{(1-u)(1-v(1-f))}{1+  \frac{(N-2)u}{2}+\frac{(N-2)(1-u)v}{3}(1-f)}$\,$\Phi_5(f)=\frac{(2-u)(1-v(1-f)) }{2+\frac{2(N-2)u}{3}+\frac{(N-2)(2-u)v}{3}(1-f)}$\,
$\Psi_1(f)=\frac{1-v(1-f)}{1+(N-1)v(1-f)}$\, $\Psi_2(f)=\frac{(1-u)(1-v(1-f))}{1+(N-1)u+(N-1)(1-u)v(1-f)}$\end{scriptsize}\\
\begin{scriptsize}$\alpha_1^{'}=\frac{1}{N-(N-1)(1-u)^2}$\, $\Phi_1^{'}(f)=\frac{(1-v(1-f))^2}{N^2-(N-1)(N-2)(1-v(1-f))^2}$\,
$\Phi_2^{'}(f)=\frac{(1-u)^3(1-v(1-f))^2}{N^2-(N-1)(N-2)(1-u)^3(1-v(1-f))^2}$\end{scriptsize}\\
\begin{scriptsize}$\Phi_3^{'}(f)=\frac{(1-u)^2(1-v(1-f))^2}{N^2-(N-1)(N-2)(1-u)^2(1-v(1-f))^2}$\,
$\Psi_1^{'}(f)=\frac{1}{N-(N-1)(1-v(1-f))^2}$\,
$\Psi_2^{'}(f)= \frac{1}{N-(N-1)(1-u)^2(1-v(1-f))^2}$\end{scriptsize}\\
\end{flushleft}
\label{table3}
\end{adjustwidth}
\end{table}

\begin{figure}[!h]
\includegraphics[width=\linewidth]{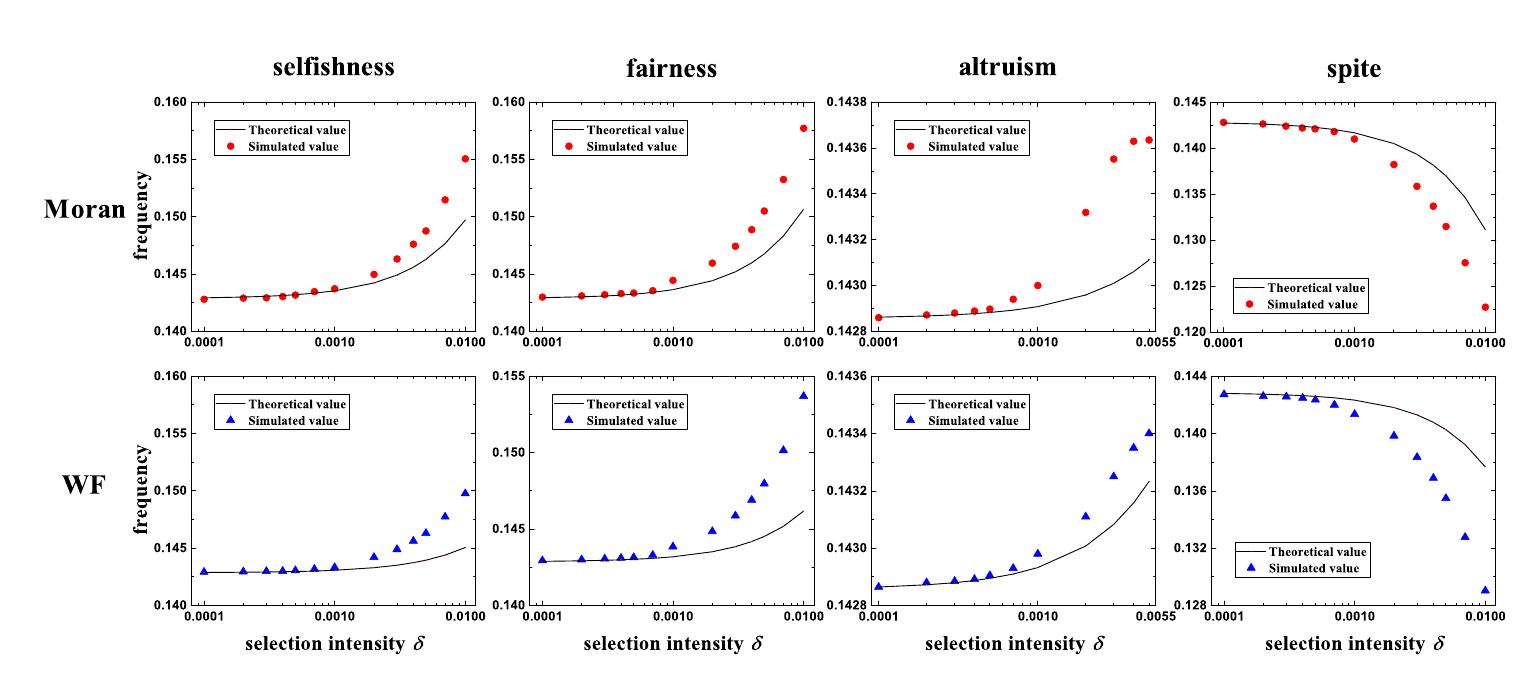}
\caption{{\bf For different} {\boldmath $N$,} {\bf comparison of} {\boldmath $f_1$, $f_2$, $f_3$,} {\bf and} {\boldmath $f_4$} {\bf over the area spanned by} {\boldmath $v$} {\bf and} {\boldmath $u$.}
For the Moran process (first row) and the Wright-Fisher process (second row),
there are two phenomena $f_2>f_1>f_3>f_4$ and $f_2>f_1>f_4>f_3$ in the small population of $N=20$; there only remains one phenomenon $ f_2>f_1>f_3>f_4$ in the moderate population of $N=80$; $f_2>f_1>f_3>f_4$ remains and there appears a new phenomenon $f_2>f_3>f_1>f_4$ in the large population of $N=660$. Parameters: $M=7$, $r=1$, and $p=0.01$.}
\label{fig3}
\end{figure}

We now focus on the effect of the population size $N$ on the seven-strategy competition in Fig~\ref{fig3}.
Irrespective of the population size, fairness ($S_2$ and $S_7$) has a higher frequency than selfishness ($S_1$), altruism ($S_3$ and $S_5$), and spite ($S_4$ and $S_6$).
Accordingly, a larger population size cannot change the dominance of fairness.
In small populations, selfishness has the second highest frequency.
Meanwhile, spite has an advantage over altruism, i.e., $f_2>f_1>f_4>f_3$, for high mutation probabilities;
the opposite, i.e., $f_2>f_1>f_3>f_4$, holds for the remaining migration probabilities ($v$) and mutation probabilities ($u$).
In moderate populations, the former disappears and the latter holds for the whole area spanned by $v$ and $u$.
In large populations, the $(v,u)$ area for the latter diminishes and is restricted to low mutation probabilities, and a new phenomenon appears in which altruism gains an advantage over selfishness, i.e., $f_2>f_3>f_1>f_4$, for the remaining values of $u$ and $v$.
In other words, the increase of the population size can raise the frequency ranking of altruism and reduce the frequency rankings of selfishness and spite.
Accordingly, increasing the population size enhances the evolution of altruism and weakens the evolution of selfishness and spite.
The above results hold for the Moran process and the Wright-Fisher process.
In small populations, the shape of the boundary of two possible phenomena varies from the Moran process to the Wright-Fisher process.
In large populations, the shape of the boundary of two possible phenomena
remains the same for these two update rules.

\begin{figure}[!h]
\includegraphics[width=\linewidth]{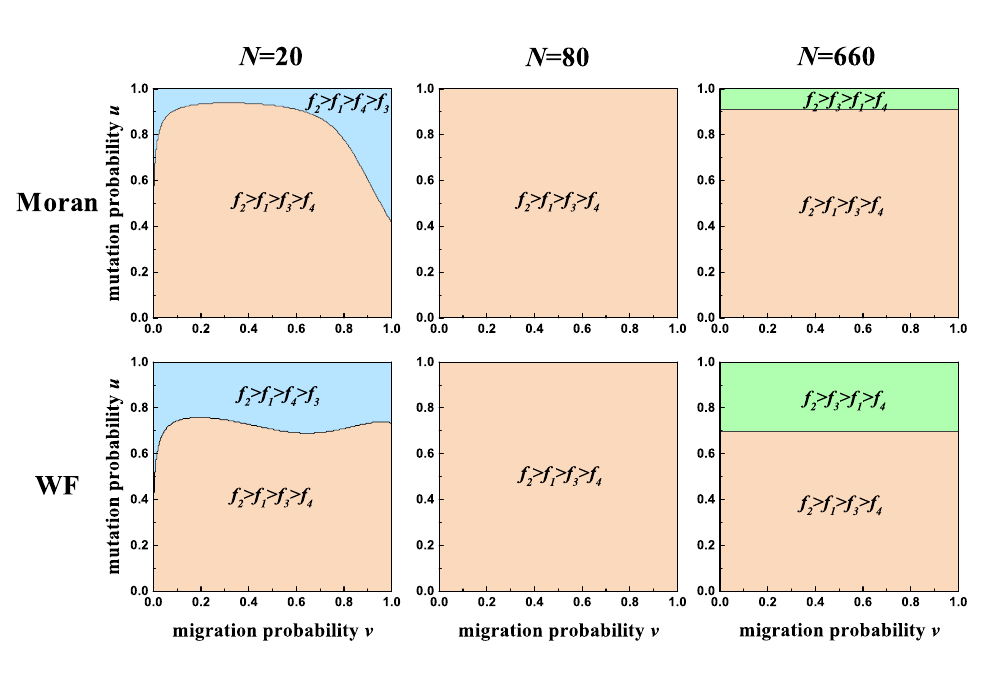}
\caption{{\bf The changing trends of} {\boldmath $f_i,i\in\{1,2,3,4\}$} {\bf with} {\boldmath $u$.}
A, C: For the Moran process with $v=0$ and the Wright-Fisher process with $v=0$, $f_1$ and $f_2$ exhibit inverted U-shaped curves with $u$, but $f_3$ and $f_4$ exhibit U-shaped curves with $u$.
B, D: For the Moran process with $v=0.05$ and the Wright-Fisher process with $v=0.05$, $f_1$, $f_2$, and $f_3$ exhibit inverted U-shaped curves with $u$, but $f_4$ exhibits a U-shaped curve with $u$.
Parameters: $N=80$, $M=7$, $r=1$, and $p=0.01$.}
\label{fig4}
\end{figure}

It is shown how mutation influences the seven-strategy competition in Fig~\ref{fig4}.
The way that mutation changes selfishness, fairness, altruism, and spite is qualitatively similar for the Moran process and the Wright-Fisher process.
In the absence of migration (the migration probability $v=0$), selfishness and fairness exhibit inverted U-shaped curves with the mutation probability $u$, but altruism and spite exhibit U-shaped curves with $u$.
A small increase of the migration probability causes altruism to change with the mutation probability $u$ from a U-shaped curve to an inverted U-shaped curve and maintains the change of selfishness, fairness, and spite with $u$.
The results for sufficiently high migration probabilities, which are not shown in Fig~\ref{fig4}, qualitatively recover to the result without migration ($v=0$).
Therefore, intermediate mutation maximizes selfishness and fairness, and minimizes spite irrespective of migration; intermediate mutation maximizes altruism for intermediate migration and minimizes altruism otherwise.

\begin{figure}[!h]
\includegraphics[width=\linewidth]{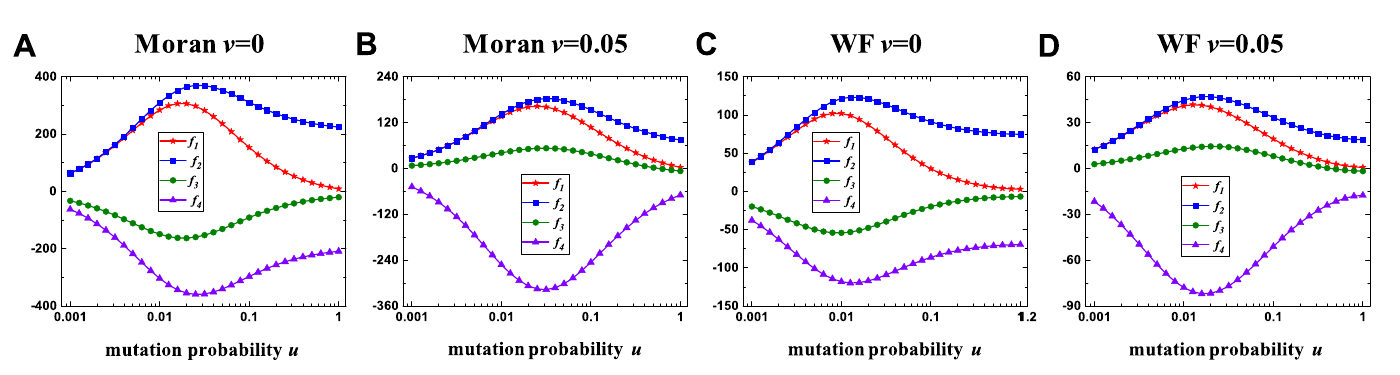}
\caption{{\bf The changing trends of} {\boldmath $f_i,i\in\{1,2,3,4\}$} {\bf with} {\boldmath $v$} {\bf for the Moran process (first row) and the Wright-Fisher process (second row) when} {\boldmath $u$} {\bf is low.}
$f_1$ (first column) and $f_2$ (second column) have decreasing trends with $v$, and $r=1$ maximizes $f_1$ and $f_2$ for all values of $v$.
The curve of $f_3$ (third column) with $v$ is inverted U-shaped, and the migration range maximizing $f_3$ goes from $r=3$ to $r=1$
with the increase of $v$.
The curve of $f_4$ with $v$ (fourth column) has an increasing trend and a small perturbation near $v=0.01$, and $r=1$ maximizes $f_4$ for very small $v$ and minimizes $f_4$ for the remaining large $v$.
When $r=1$, the curve of $f_i,i\in\{1,2,3,4\}$ with $v$ has a small perturbation near $v=1$ for the Wright-Fisher process, but not for the Moran process. Parameters: $N=80$, $M=7$, $u=0.01$, and $p=0.01$.}
\label{fig5}
\end{figure}

\begin{figure}[!h]
\includegraphics[width=\linewidth]{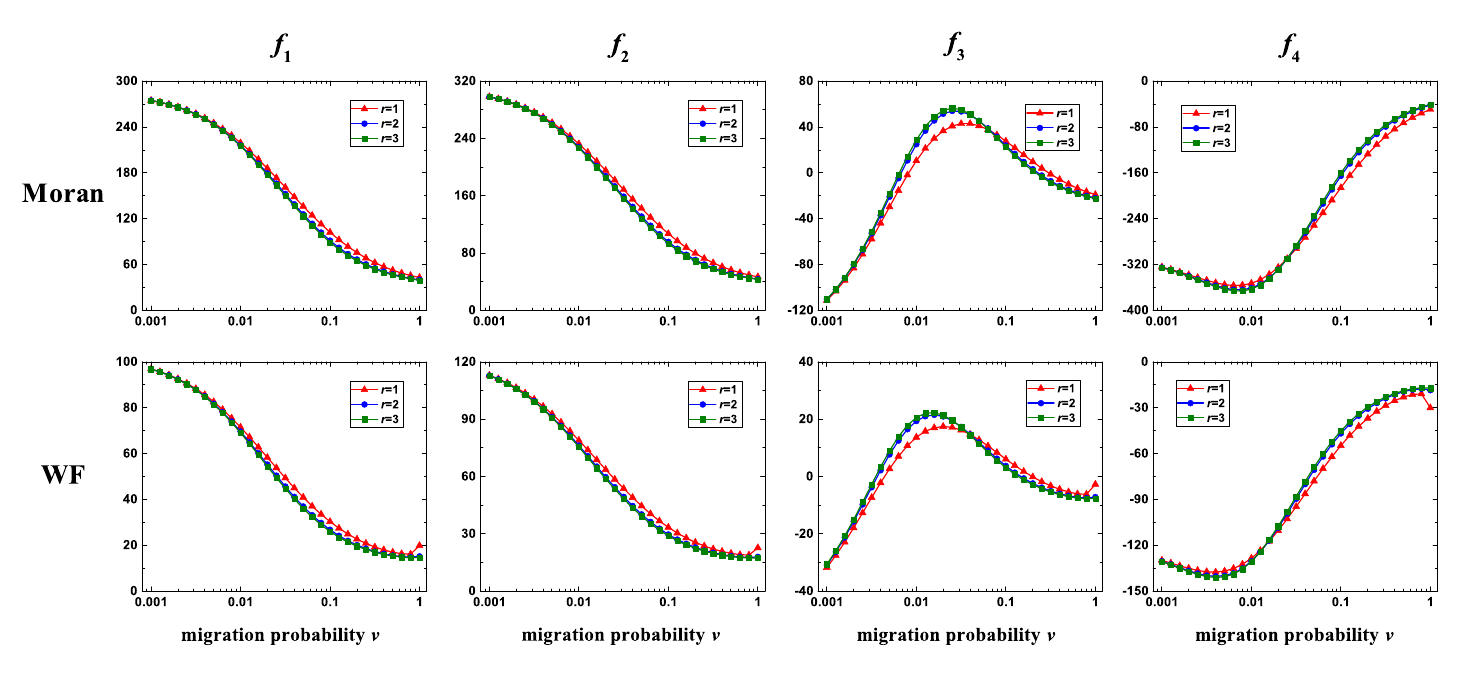}
\caption{{\bf The changing trends of} {\boldmath $f_i,i\in\{1,2,3,4\}$} {\bf with} {\boldmath $v$} {\bf for the Moran process (first row) and the Wright-Fisher process (second row) when} {\boldmath $u$} {\bf is high.}
$f_1$ (first column) and $f_2$ (second column) decrease with $v$, and $r=1$ maximizes $f_1$ and $f_2$ for all values of $v$.
$f_3$ (third column) and $f_4$ (fourth column) increases with $v$, and $r=1$ minimizes $f_3$ and $f_4$ for all values of $v$. Parameters: $N=80$, $M=7$, $u=1$, and $p=0.01$.}
\label{fig6}
\end{figure}

We now investigate the effect of migration (probability and range) on the seven-strategy competition when the mutation probability $u$ is low (Fig~\ref{fig5}) and high (Fig~\ref{fig6}).
Migration changes selfishness and fairness qualitatively similarly: they both have decreasing   trends with the migration probability $v$, and the smallest migration range $r=1$ maximizes both of them.
These results hold for all mutation probabilities.
Compared with the case without migration, the existence of migration decreases selfishness and fairness, and thus inhibits the evolution of selfishness and fairness.
The way that migration changes altruism varies from low to high mutation probabilities:
when the mutation probability $u$ is low, there exists a moderate migration probability $v$ which maximizes altruism, and the migration range maximizing altruism is from the largest to the smallest range with the increase of $v$;
when $u$ is high, altruism increases with $v$, and the smallest migration range minimizes altruism.
Compared with the case without migration, the existence of migration increases altruism, and thus promotes the evolution of altruism.
The way that migration changes spite also varies from low to high mutation probabilities:
when the mutation probability $u$ is low, the curve of spite with the migration probability $v$ has an increasing trend and a small perturbation near $v=0.01$, and the smallest migration range maximizes spite for very small migration probabilities and minimizes spite for the remaining majority of migration probabilities;
when $u$ is high, spite increases with $v$, and the smallest migration range minimizes spite.
Compared with the case without migration, sufficient migration ($v$ is not too low when $u$ is low) increases spite, and thus promotes the evolution of spite.
The above results are appropriate for the Wright-Fisher process and the Moran process.
These two update rules have a qualitative difference
for the smallest migration range and small mutation probabilities: the curves of selfishness, fairness, altruism, and spite with the migration probability $v$ have small perturbations at $v=1$ for the latter but not for the former.

\section*{Conclusion}
We have studied the effect of spite and altruism on the evolution of fairness in an evolutionary model using the simplified version of the UG with seven representative discrete strategies.
A simplified version of the UG with four discrete strategies has been used to investigate the effect of spite on the evolution of fairness in a previous evolutionary model~\cite{Forber2014}.
The previous model assumed two kinds of offers
(low and fair) and two kinds of veto power (accept any and reject low), and did not consider high-offer rejections which have been found in the form of non-monotonic rejections (rejecting offers that are too high and too low) in experiments~\cite{Bahry2006,Hennig-Schmidt2008}.
Based on the previous model~\cite{Forber2014}, we have added a high offer and a non-monotonic rejection.
Our model includes the four strategies in the previous model (separately represent selfishness, fairness, altruism, and spite) and has three new strategies (separately  represent altruism, spite, and fairness).
Moreover, the analysis method in our study is different from that in the pervious study~\cite{Forber2014}.
The previous study has only investigated the four-strategy competition in infinite populations, and has implicitly demonstrated that spite promotes the evolution of fairness under certain conditions.
Our study has used strategy intervention in finite populations to explicitly study how spite influences the evolution of fairness: we start with the competition between a selfish strategy and a fair strategy; we then add five strategies to the competition in sequence, and our study goes from the three-strategy competition to the seven-strategy competition.
In the two-strategy competition, selfishness competes equally with fairness.
The addition of altruism leads to the advantage of selfishness over fairness,
and this advantage can be offset by the further addition of spite.
Accordingly, we have explicitly shown that altruism inhibits the evolution of fairness, whereas spite promotes the evolution of fairness.
Fairness first gains an advantage over selfishness when the fair strategy with the non-monotonic rejection is added, and thus we have found that the non-monotonic rejection can cause fairness to overcome selfishness, which cannot happen without the high-offer rejection.
It is worth mentioning that the four-strategy competition of our model corresponds to the finite-population version of the previous model~\cite{Forber2014}.

Most previous studies about the UG, including the above-mentioned literature~\cite{Forber2014},
have neglected the role of population finiteness in the evolution of fairness.
Traditional deterministic evolutionary dynamics~\cite{Nowak2000} has shown that fairness cannot evolve in infinite populations without other mechanisms.
Until recently, a stochastic evolutionary model~\cite{Rand2013} has demonstrated that fairness can evolve in finite populations without any other mechanisms, meaning that the finiteness of the population matters in the evolution of fairness.
In this paper, structured populations of finite size satisfying the Tarnita-$\sigma$ condition~\cite{Tarnita2011} have been used to study the simplified version of the UG.
Particularly for the group-structured population together with the Moran process or the Wright-Fisher process, we have obtained the concrete Tarnita-$\sigma$ condition (without unknown parameters), based on the previous calculation~\cite{Zhang2016b}, which has only been used to analyze the multiple-strategy competition in general models.
By using the concrete Tarnita-$\sigma$ condition, we have studied the effect of the population size on the seven-strategy competition.
For the Moran process and the Wright-Fisher process, a larger population size cannot change the dominance of fairness, but it enhances the evolution of altruism and weakens the evolution of selfishness and spite.

The effect of migration on the evolution of fairness has been previously studied by agent-based simulations~\cite{Wang2015}.
Here, we have given the analytic results about how migration and mutation influence the seven-strategy competition.
The Moran process and the Wright-Fisher process have the following qualitatively similar results.
Intermediate mutation maximizes selfishness and fairness, and minimizes spite irrespective of migration; intermediate mutation maximizes altruism for intermediate migration and minimizes altruism otherwise.
The existence of migration inhibits the evolution of selfishness and fairness, and promotes the evolution of altruism; sufficient migration promotes the evolution of spite.
The single qualitatively different result between the Moran process and the Wright-Fisher process lies at the smallest migration range and small mutation probabilities: the curves of selfishness, fairness, altruism, and spite with the migration probability $v$ have small perturbations at $v=1$ for the latter but not for the former.

\section*{Model and methods}
\subsection*{Model}
\begin{figure}[h!]
\includegraphics[width=\linewidth]{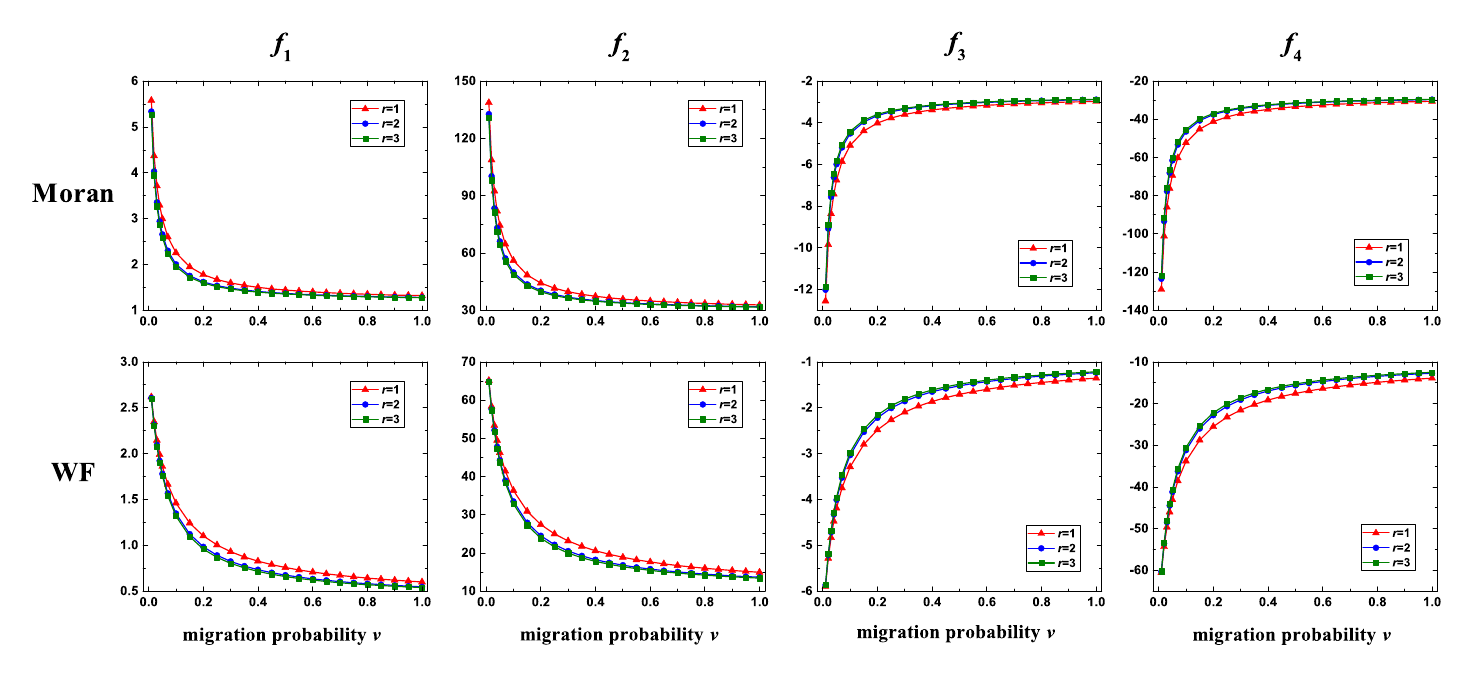}
\caption{{\bf Model schematic.}
(A) The simplified version of the UG with seven representative discrete strategies.
(B) Migration of the range $r$ when seven groups are located in a circle.}
\label{fig7}
\end{figure}

In the UG, the proposer has to divide a certain amount of money, say $1$, with the responder who can accept or reject the split.
If the responder accepts the split, the money is shared accordingly; if not, both individuals remain empty handed.
We focus on a simplified version of the UG in Fig~\ref{fig7}.
Proposers have three kinds of offers: fair ($0.5$), low ($p>0, p\to 0$), and high ($0.5+p$), in which the first one is an equal offer for the proposer and the responder and the latter two are unequal offers.
Experiments which investigated the responder behavior~\cite{Bahry2006,Hennig-Schmidt2008}
have displayed that many responders use non-monotonic rejections (rejecting offers that are too high and too low).
Accordingly, we here assume that responders have three kinds of veto power: accept any, reject low, and reject unequal (reject low and high).
Seven representative discrete strategies, each of which denotes what choice to make as a proposer and what choice to make as a responder, will be used: $S_1=(p, \mbox{accept any})$, $S_2=(0.5, \mbox{reject low})$, $S_3=(0.5, \mbox{accept any})$, $S_4=(p, \mbox{reject low})$, $S_5=(0.5+p, \mbox{accept any})$, $S_6=(p, \mbox{reject unequal})$, and $S_7=(0.5, \mbox{reject unequal})$.

The individual using $S_1$ only cares about maximizing his own payoff, and thus $S_1$ represents selfishness.
The individual using $S_2$ or $S_7$ is willing to sacrifice his own payoff to pursue fairness (the proposer gives up his priority and plays fair to the opponent, and the responder punishes the low offer or the unequal offers), and thus $S_2$ and $S_7$ represent fairness.
The individual using $S_3$ or $S_5$ is kind to his opponent using any strategies (the proposer volunteers to give the opponent a non-low offer, and the responder allows the opponent to obtain a higher payoff than himself), and thus $S_3$ and $S_5$ represent altruism.
The individual using $S_4$ or $S_6$ is concerned with his own relative standing (the proposer is always trying to get a higher payoff than the opponent, and the responder will not leave his own payoff below his opponent's), and thus
$S_4$ and $S_6$ represent spite.

We will first consider the population structure (determines whom an individual interacts with) and the update rule (determines how individuals compete for reproduction) satisfying the Tarnita-$\sigma$ condition.
Any one given interaction is comprised of two games, in which two individuals play the roles of proposer and responder alternately.
The payoff matrix for the simplified version of the UG is shown in Table~\ref{table1}.
All interactions accumulate the payoff of individual $i$, $p_i$, and further his fitness, $f_i=1+\delta p_i$.
There may occur mutation during reproduction:
mutation happens to an offspring with probability $u$, and then he equi-probably chooses one of all possible strategies; otherwise (with probability $1-u$), the offspring inherits the strategy of his parent.

\begin{table}[!ht]
\centering
\caption{{\bf The payoff matrix for the simplified version of the UG.}}
\fboxsep=100pt
\begin{tabular}{|l+l|l|l|l|l|l|l|}
\hline
& {\boldmath $S_1$} &{\boldmath $S_2$}&{\boldmath $S_3$}&{\boldmath $S_4$}&{\boldmath $S_5$}&{\boldmath $S_6$}&{\boldmath $S_7$}\\
\thickhline
{\boldmath $S_1$}({\boldmath $p$}, {\bf accept any}) & 1 & 1/2 & 3/2-$p$ &  $p$ & 3/2 &  $p$&1/2\\ \hline
{\boldmath $S_2$}({\boldmath $0.5$}, {\bf reject low}) & 1/2 & 1 & 1 &  1/2 & 1+$p$ &  1/2 &  1\\ \hline
{\boldmath $S_3$}({\boldmath $0.5$}, {\bf accept any}) & 1/2+$p$ & 1 & 1 &  1/2+$p$ & 1+$p$ &1/2+$p$&1\\ \hline
{\boldmath $S_4$}({\boldmath $p$}, {\bf reject low}) & 1-$p$ & 1/2 & 3/2-$p$ &  0 & 3/2 & 0 &1/2\\ \hline
{\boldmath $S_5$}({\boldmath $0.5+p$}, {\bf accept any}) & 1/2 & 1-$p$ & 1-$p$ &  1/2 & 1 &  $p$ &  1/2\\ \hline
{\boldmath $S_6$}({\boldmath $p$}, {\bf reject unequal}) & 1-$p$ & 1/2& 3/2-$p$ &0 & 1-$p$ &0& 1/2\\ \hline
{\boldmath $S_7$}({\boldmath $0.5$}, {\bf reject unequal}) & 1/2 & 1 & 1 &  1/2 & 1/2 &  1/2 &  1\\ \hline
\end{tabular}
\label{table1}
\end{table}

We will then consider two particular models, the Moran process and the Wright-Fisher process in the group-structured population.
The whole population of size $N$ is distributed over $M$ groups which are located in a circle, and an individual only interacts with all others of the same group.
In the Moran process, all individuals of the population compete to reproduce one offspring proportional to their fitness, and then one individual is equi-probably chosen from the whole population to die.
In the Wright-Fisher process, all individuals of the population compete to reproduce $N$ (population size) offspring proportional to their fitness, and the whole population is replaced by all the newborn offspring.
Besides mutation, migration is also introduced into our model: with probability $1-v$, the offspring remains in his parent's group; with probability $v$, he moves to a new group according to the migration pattern of the range $r$ in Fig~\ref{fig7}.
The migration range $r$ means that all possible displacements generated by a single-step migration form the set $\Omega(r)=\{1,2,\cdots,r\}$ and all elements of $\Omega(r)$ are performed equi-probably.

\subsection*{Methods}
Let $F_k=\Gamma_1(a_{kk}-\overline{a}_{**})+\Gamma_2
(\overline{a}_{k*}-\overline{a}_{*k})+\Gamma_3(\overline{a}_{k*}-\overline{a})$.
Under weak selection, the stationary frequency of $S_i$ is greater than that of $S_j$ if and only if $F_i$ is greater than $F_j$.
When selfishness, fairness, altruism, or spite
is exhibited by a single strategy, we assume $f_i=F_i$ with $i\in\{1,2,3,4\}$.
When fairness, altruism, or spite is exhibited by two strategies, we assume $f_2=(F_2+F_7)/2$, $f_3=(F_3+F_5)/2$, or $f_4=(F_4+F_6)/2$.
This is because all possible strategies have similar frequencies under weak selection and
our assumption can guarantee that the comparison among selfishness, fairness, altruism, and spite proceeds on the same scale.
The comparison is based on $f_1$, $f_2$, $f_3$, and $f_4$ in Table~\ref{table2}.
For example, selfishness has an advantage over fairness if and only if $f_1>f_2$, the reverse holds if and only if $f_1<f_2$, and they compete equally if and only if $f_1=f_2$.

\begin{table}[!ht]
\begin{adjustwidth}{-2.25in}{0in} 
\centering
\caption{
{\boldmath $f_1$, $f_2$, $f_3$,} {\bf and} {\boldmath $f_4$}.}
\begin{tabular}{|l+l|l|l|l|}
	\hline
		&	& {\boldmath $f_2=F_2$} {\bf or}  &{\boldmath $f_3=F_3$} {\bf or} & {\boldmath $f_4=F_4$} {\bf or} \\
	   &\multirow{-2}*{{\boldmath $f_1=F_1$}} &{\boldmath $f_2=(F_2 + F_7)/2$}& {\boldmath $f_3=(F_3 + F_5)/2$} & {\boldmath $ f_4=(F_4 + F_6)/2$}\\\thickhline
				
	{\bf three-strategy} & &   &  &  \\
    {\bf competition}    &\multirow{-2}*{ $\frac{3( 1 - 2p ){\Gamma _2} + ( 1 - 3p )\Gamma_3}{9}$} & \multirow{-2}*{$- \frac{\Gamma _3 }{18}$}& \multirow{-2}*{$ - \frac{6( 1 - 2p)\Gamma _2 + ( 1 - 6p)\Gamma _3}{18}$} & \\	\hline	
	{\bf four-strategy} &	&  & &  \\
	{\bf competition}   &\multirow{-2}*{$\frac{\Gamma _1}{4}$} &\multirow{-2}*{$\frac{\Gamma _1}{4}$}&\multirow{-2}*{ $\frac{\Gamma _1 - 2( 1 - 2p )\Gamma _2 + 2p\Gamma _3}{4}$} & \multirow{-2}*{ $\frac{- 3\Gamma _1 + 2( 1 - 2p)\Gamma _2 - 2p\Gamma _3}{4}$}\\	\hline
	{\bf five-strategy}	&	&   &  &  \\
	{\bf competition}     &\multirow{-2}*{$\frac{10\Gamma _1 + 10\Gamma _2 + 3\Gamma _3}{50}$} & \multirow{-2}*{$\frac{5\Gamma _1 + 10p\Gamma _2 - ( 1 - 5p )\Gamma _3}{25}$} &\multirow{-2}*{$\frac{10\Gamma _1 - 10( 2 - p )\Gamma _2 - ( 2 - 50p)\Gamma _3}{50}$} &\multirow{-2}*{$\frac{ - 40\Gamma _1 + 10( 3 - 4p )\Gamma _2 + ( 3 - 20p )\Gamma _3}{50}$}\\\hline
	 {\bf six-strategy} &	&  & &  \\
	 {\bf competition}	&\multirow{-2}*{$\frac{12\Gamma _1 + 12p\Gamma _2 + ( 1 + 6p )\Gamma _3}{36}$} &\multirow{-2}*{$\frac{12\Gamma _1 + 12p\Gamma _2+ ( 1 + 6p)\Gamma _3}{36}$} & \multirow{-2}*{ $\frac{24\Gamma _1 - 36( 1 - p )\Gamma _2 - ( 1 - 18p )\Gamma _3}{72}$}&\multirow{-2}*{ $\frac{- 48\Gamma _1 + ( 36 - 60p )\Gamma _2 - ( 1 + 30p )\Gamma _3}{72}$}\\\hline
	{\bf seven-strategy}	&	&   &  &  \\
	{\bf competition} &\multirow{-2}*{$\frac{2\Gamma _1 + 2p\Gamma _2 + p\Gamma _3}{7}$} & \multirow{-2}*{$\frac{8\Gamma _1 + 4p\Gamma _2 + ( 1 + 2p )\Gamma _3}{28}$} &\multirow{-2}*{$\frac{4\Gamma _1 - 6( 1 - p )\Gamma _2 + 3p\Gamma _3}{14}$} &\multirow{-2}*{$\frac{ - 20\Gamma _1 + 4( 3 - 5p )\Gamma _2 - ( 1 + 10p )\Gamma _3}{28}$}\\		\hline
\end{tabular}
\begin{flushleft} Three-strategy competition is the competition of $s_1, s_2, s_3$, four-strategy competition is the competition of $s_1, s_2, s_3, s_4$, five-strategy competition is the competition of $s_1, s_2, s_3, s_4, s_5$, six-strategy competition is the competition of $s_1, s_2, s_3, s_4, s_5, s_6$, and seven-strategy competition is the competition of $s_1, s_2, s_3, s_4, s_5, s_6, s_7$.
\end{flushleft}
\label{table2}
\end{adjustwidth}
\end{table}

\section*{Acknowledgments}
Y. Z. is grateful for support by China Postdoctoral Science Foundation (No. 2015M580989) and National Natural Science Foundation
of China (No. 61603036). F. F. acknowledges generous support from the Dartmouth Faculty Startup Fund, Walter \& Constance Burke Research Initiation Award, NIH (No. C16A12652), and DARPA (No. D17PC00002-002).

\nolinenumbers

%
%
%

\end{document}